\documentclass[12pt]{article}
\usepackage{graphicx}
\usepackage{amssymb}

\newcommand{\beq}{\begin{equation}}
\newcommand{\eeq}{\end{equation}}
\newcommand{\bq}{\begin{quotation}}
\newcommand{\eq}{\end{quotation}}
\newcommand{\bc}{\begin{center}}
\newcommand{\ec}{\end{center}}

\begin{document}

\title{{\vspace{0cm}
\addtocounter{footnote}{2}
\sc A critical note on time in \\
the multiverse}}

\author{
\addtocounter{footnote}{-2}
{\sc S.E. Rugh\footnote{Symposion, 
`The Socrates Spirit', Section for Philosophy and the Foundations of Physics,
Helleb\ae kgade 27, Copenhagen N, Denmark
({\em e-mail: rugh@symposion.dk)}} 
\addtocounter{footnote}{5}
and  H.
Zinkernagel\footnote{Department of Philosophy I, Granada University, 18071
Granada, Spain ({\em e-mail: zink@ugr.es}).}} }
\date{}

\maketitle

\begin{abstract}
\noindent
In recent analyses of standard, single-universe, cosmology, it was pointed out that specific assumptions regarding the distribution and motion of matter must be made in order to set up the cosmological standard model with a global time parameter. Relying on these results, we critically examine the notion of time
in the multiverse -- and in particular the idea that some parts of the multiverse are older than others. By focusing on the most elaborated multiverse proposal in cosmology, the inflationary multiverse, we identify three problems for establishing a physically well-defined notion of global time; a quantum problem, a collision problem and a fractal problem. The quantum problem -- and the closely related ``cosmic measurement problem" -- may even undermine the idea that parts of the multiverse causally and temporally precede our universe. 
\end{abstract}

\section{Introduction}

The idea of a ‘multiverse’ has recently become quite popular in modern cosmology.  According to some multiverse scenarios, based e.g. on so-called chaotic inflation, our universe is supposed to be just one inflating bubble in a much bigger and older multiverse  – with each component expanding differently and having different physical laws (see e.g. Linde 2004 and Guth 2007). In this and related versions, the multiverse thus purports to reject the common wisdom regarding modern cosmology according to which asking what was before the ‘big bang’ is considered as meaningless as asking what is north of the North Pole, see e.g. Hawking (1989, p. 69).

While the multiverse idea has been widely discussed (and criticized) e.g. in connection with its apparent lack of empirical testability (see e.g. Carr and Ellis 2008) very few studies have addressed the more conceptual problems facing the notion of a multiverse in cosmology.\footnote{For a brief review and references to some of these problems, see Zinkernagel (2011).}
In this paper, we want to explore a little discussed conceptual question about the multiverse: Does it include a sensible {\em{notion of time}} which allows us to speculate that it is not only much bigger but also much {\em{older}}  than our (local) universe? The answer to this question will obviously depend both on what kind of multiverse is contemplated, and on how time is (or could be) conceived in the specific multiverse proposal. In any case, the investigation of the question is likely to contribute to a clarification of the conceptual foundation of cosmology.

The outline of the paper is as follows. We first review some earlier work which shows that a relationist understanding of time (an interdependence between time, matter and motion) is essential to the standard notion of cosmic time. Armed with this clarification, we discuss possible ways to understand the claim that there are older patches (than our universe) in the multiverse. After that we discuss the most worked out version of the multiverse arising from the theory of inflation and question whether the notion of time in this theory is applicable as a multiverse time. In the closing section, we offer a few brief comments on other multiverse scenarios and note that these are likely to be even worse off with regard to time than the inflationary multiverse.\footnote{We explore the notion of time in both the universe and the multiverse in more detail in Rugh and Zinkernagel (2013).}

\section{Time in standard (single-universe) cosmology}

In our earlier work we have defended a version of relationism which affirms that time is necessarily associated with physical processes. More specifically, we argue in favour of a `time-clock' relation which asserts that time, in order to have a physical basis, must be understood in relation to physical processes which act as `cores' of clocks (Rugh and Zinkernagel 2005, 2009, see also Zinkernagel 2008). In the cosmological context, the time-clock relation implies that a necessary physical condition for {\em interpreting}  the $t$ parameter of the standard Friedmann-Lema\^{\i}tre-Robertson-Walker (FLRW) model as cosmic time in some `epoch' of the universe is the (at least possible) existence of a physical process which can function as a core of a clock in the `epoch' in question.

There is a more direct route to relationism in cosmology which is independent of the mentioned time-clock-relation (even if in conformity with it). In this regard, we discuss in Rugh and Zinkernagel (2011) how the very set-up of the FLRW model with a global time is closely linked to the motion, distribution and properties of cosmic matter. In the following, we briefly review some key points of this discussion which are necessary components of our analysis of time in the multiverse.

In relativity theory time depends on the choice of reference frame. Since, for a universe, a reference frame cannot be given from the outside, such a frame has to be ``built up from within", that is, in terms of the (material) constituents within the universe. It is often assumed that the FLRW model 
may be derived just from the cosmological principle. This
principle states that the universe is spatially homogeneous and isotropic (on large scales). It is much less well 
known that another assumption, called Weyl's principle, is necessary in order to arrive at the FLRW model and, in particular, its cosmic time parameter.  Whereas the cosmological principle imposes constraints on the {\em {distribution}} of the matter content of the universe, Weyl's principle imposes constraints on the {\em {motion}} of the matter content. Weyl's principle (from 1923) says that the matter content is so {\em{well behaved}} that a reference frame can be built up from it: 

\begin{quotation}
\noindent
Weyl's principle (in a general form): The world lines of `fundamental
particles' form a spacetime-filling family of non-intersecting geodesics (a congruence of geodesic world lines).
\end{quotation}

The importance of Weyl's principle is that it provides a reference frame which is physically based
on an expanding `substratum' of `fundamental particles' (e.g. galaxies or clusters of galaxies).
In particular, if the (non-crossing)
geodesic world lines are required to be orthogonal to a series of space-like hypersurfaces, a comoving reference frame is defined in which constant spatial coordinates are ``carried by" the fundamental particles (see figure 1 in section \ref{defined}). The time coordinate is a cosmic time which labels the series of hypersurfaces, and which may be taken as the proper time along any of the particle world lines. We note that the congruence of world lines is essential to the standard cosmological model since the symmetry constraints of homogeneity and isotropy are imposed w.r.t. such a congruence (see e.g. Ellis 1999). Thus, Weyl's principle is {\em{a precondition}} for the cosmological principle; the former can be satisfied without the latter being satisfied but not vice versa.

\subsection{Is the Weyl principle (always) satisfied in our universe?}
\label{FLRW satisfied}

There are several possible problems which may arise with the Weyl principle. First, there is the question of whether particle trajectories are always well-defined (at all times in cosmic history). Second, whether -- if such well-defined trajectories cross -- a suitable averaging procedure exists for smoothing out these crossings. As regards the latter problem, it is clear that Weyl's principle cannot hold for ordinary galaxies as they indeed may (and do) collide. Likewise with the more fundamental constituents in earlier phases of the universe.
Thus the fundamental world lines in the Weyl principle must be 
some `average world lines' associated with the average motion
of the fundamental particles over some coarse-grained scales (in order to ``smooth out" any crossings).\footnote{There exist observationally based claims (e.g. Labini et al. 2009) that the matter distribution is not homogeneous but instead fractal at intermediate scales at least up to distances of the order $\sim 100$ Mpc. If this fractality extended to arbitrary large distance scales, there would be no scale above which collisions could be averaged out. Moreover, there would be `holes' on all scales so no set of `average world lines' could fill space-time (implying that no congruence could be formed), and also a homogeneous universe could not be recovered. Thus, both the Weyl principle and the cosmological principle -- even in their `coarse grained' versions -- would be undermined  (see Rugh and Zinkernagel 2013).} 

Regarding the first problem of whether particle trajectories can at all be identified, the starting point is that the Weyl principle refers to a 
non-crossing family of (fluid or particle) world lines. The notion of such lines refers to classical, or classicalized, particle-like behavior of the material constituents. This makes it difficult to even formulate the Weyl  principle (let alone decide whether it is satisfied) if some period in cosmic history is reached (in a backward extrapolation from now) where the `fundamental particles' are to be described by wave-functions $\psi (x,t)$ referring to entangled quantum constituents. What is a `world line' or a `particle trajectory' then? Unless one can specify a clear meaning of non-intersecting trajectories in a contemplated quantum `epoch', it would seem that the very notion of cosmic time, and hence the notion of `very early universe' is compromised. This last problem of identifying a Weyl substratum within a quantum description arises most clearly on a ``quantum fundamentalist" view according to which the material constituents of the universe could be described {\em exclusively} in terms of quantum theory at some early stage of the universe.
As noted in Rugh and Zinkernagel (2011), there is still no good answer to what may be called the ``cosmic measurement problem" (how to get classical structures from quantum constituents in a cosmological context), not least because it is highly questionable whether decoherence is sufficient to explain the building up of a Weyl substratum.

\section{Time in the multiverse?}

With the above considerations concerning time in standard cosmology, we are now ready to tackle the question of time in the multiverse. More specifically, we ask whether -- and under which conditions -- one is justified in contemplating the idea that some parts of the multiverse are older than ours. There seem to be at least two relevant ways to establish the possibility of older patches or bubbles:

\begin{enumerate}
\item Define some sort of a `multiverse' (or `supercosmic') time for the multiverse which gives a definite time ordering of the patches (as in figure 2 below).

\item If this cannot be done, then try to extrapolate our `local' cosmic time concept back through our `local' big bang.\footnote{A third and related possibility, which we shall discuss below, is to use proper time -- or at least a time order -- (associated with a single world line) to extrapolate backwards even in cases where no `local' cosmic time can be defined.} 
\end{enumerate}

Either way, the overall conclusion from section 2 is that time is relational. Thus, there is no freely flowing absolute and universal background time parameter so both multiverse- and (the extrapolation of a) cosmic time need to be grounded in the behavior of the constituents within the multiverse and the universe respectively.

The notion of a `multiverse' covers a great many possibilities (see e.g. Carr 2007). In order to address something relatively well-defined we shall in this 
short note restrict ourselves to consider some particular case studies  of inflationary multiverse models which, in our assessment, seem to be models (1) in which the model builders to some degree reflect upon -- or even attempt to provide a physical underpinning for -- the time concepts employed; and (2) which are investigated and developed to a degree that they have entered the contemporary standard literature on cosmology with some claims of observational testability.\footnote{Note that there are other proposals for older structures than our universe -- e.g. cyclic universes (``temporal" multiverses). Some of these have been discussed (and their time concept criticized) in Zinkernagel (2008).} The basic idea of the inflationary multiverse is that of a background (inflating) de Sitter space in which local bubble universes (where inflation quickly comes to an end in thermalization and particle production) continuously form (see figure 2). In its simplest version, the inflationary multiverse is driven by a single scalar field $\varphi$ -- the inflaton (which, at present, is unrelated to any known particle physics), see e.g. Linde (2004).

\subsection{Can a multiverse time be defined?}
\label{defined}

\begin{figure}[ht]
\begin{minipage}[b]{0.5\linewidth}
\centering
\includegraphics[scale=.6]{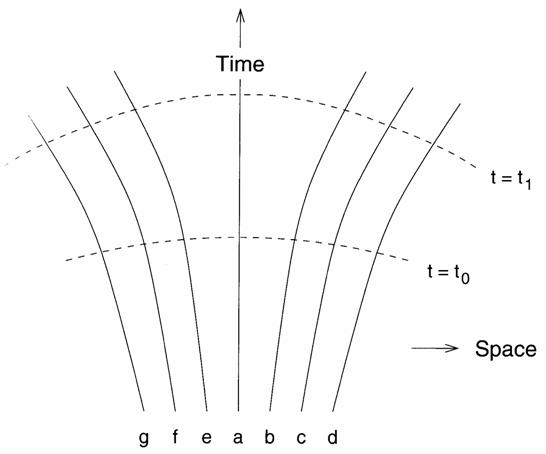}
\caption{{\small An idealized  ``Weyl substratum". The particle (e.g. galaxy) trajectories form a congruence in an approximation where galaxies are seen as space-time filling particles of a fluid. Figure from Narlikar (2002).}}
\label{fig:figure1}
\end{minipage}
\hspace{0.5cm}
\begin{minipage}[b]{0.5\linewidth}
\centering
\includegraphics[scale=.6]{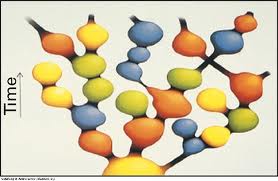}
\caption{{\small A multiverse consisting of bubble universes arising from the chaotic inflation model with a suggestive global multiverse ``time" axis indicated on the left.
 Figure from Linde (1998).}}
\label{Linde}
\end{minipage}
\end{figure}

In this paper a main question concerns whether the Weyl principle is satisfied in the multiverse. To motivate an initial doubt, consider figure \ref{Linde}, in which there does not seem to be a multiverse with patches or bubbles obeying the Weyl principle (a similar figure suggesting a multiverse time can be found in Guth 2007). Thus, there is no immediate physical basis for a multiverse time (indicated in the figure) which could order the patches.\footnote{The colours in the (original version of the) figure represent different effective physical laws (or constants) in the different bubble universes. This corresponds to a more complicated multiverse model -- with various scalar fields -- than the one  discussed below (which has only one scalar field $\varphi$). In our view, however, this complication does not change the discussion to follow.}

By making the analogy between the idealized Weyl substratum (a congruence of e.g. galaxy world-lines) in our universe (figure 1) and the picture of an infiltrated network of bubbles in the realm of the inflationary multiverse (figure 2), it is assumed that the bubble universes somehow play the role of the substratum. The only alternative will be to assume that this substratum is constructed from the part of the $\varphi$-field outside the bubbles. Either way, we see trouble. In the first case, because the bubbles collide. In the second case because it is hard to construct (Weyl) trajectories from the $\varphi$-field, e.g. due to quantum effects (see below).

The need to satisfy the Weyl principle does seem to be recognized in the multiverse literature. Thus, Vanchurin et al (2000) notes that ``an inflating universe can be locally [within a bubble] described using the synchronous coordinates $ds^2 = d\tau ^2 - a^2({\bf{x}}, \tau)d{\bf{x}}^2$". They continue:

\begin{quotation}
\noindent
The lines of {{\bf x}} = const in this [synchronous] metric are timelike geodesics corresponding to the world lines of co-moving observers,
and the coordinate system is well defined as long as the
geodesics do not cross. This will start happening only after
thermalization, when matter in some regions will start collapsing
as a result of gravitational instability. Hence, the synchronous
coordinates can be extended to the future well
into the thermalized region.
\end{quotation}

This amounts to the claim that there is a Weyl substratum (a $\varphi$-field) which allows us to set up a synchronous coordinate system within a bubble (and it is within a bubble that thermalization and particle production occur). A similar construction seems to be applied in the global case, i.e. for the whole multiverse. Thus, Guth (2007, p. 6820) remarks that one can construct ``a Robertson-Walker coordinate system while the model universe is still in the false vacuum (de Sitter) phase, before any pocket universes have formed. One can then propagate this coordinate system forward with a synchronous gauge condition".
However, even if the importance of the Weyl principle is implicitly recognized by proponents of the inflationary multiverse, we see three step-wise related problems for this principle to be satisfied (and hence for a multiverse time to be physically underpinned):

\begin{enumerate}
\item Are there well-defined trajectories in the multiverse?

\item If there are well-defined trajectories, are they non-crossing? 

\item If they cross, can such crossing trajectories be ``averaged out"?
\end{enumerate}

We elaborate a bit on these questions in the following three subsections, and as we shall see, the answers to them may well be no, no and no. This is due to what one could call, respectively, the quantum problem, the collision problem and the fractal problem.

\subsubsection*{1. Are there well-defined trajectories
in the multiverse?}

\noindent
As mentioned in subsection \ref{FLRW satisfied}, the assumption of a quantum nature of the material (or otherwise) constituents of the universe makes it hard  (or impossible) to associate these with well-defined particle trajectories. And during inflation the only relevant constituent of the universe is taken to be the inflaton field $\varphi$ which -- in the last analysis -- is a quantum field. While the quantum-classical transition from quantum fluctuations to classical density perturbations has been widely discussed (even if not critically scrutinized, for an exception see e.g. Sudarsky 2011), this point -- that the $\varphi$ field itself is a quantum field -- is easily overlooked. For instance Linde writes after describing the basic mechanism in chaotic inflation (the most simple inflation model) which ends in the oscillations of the scalar field near the minimum of its potential (p. 130 in Carr 2007): ``As any rapidly oscillating {{\em classical}} field, it loses its energy by creating pairs of elementary particles" (our emphasis). Despite the wording, this is not a reconceptualization of the whole edifice of classical field theory! Linde is, of course, well aware that it requires  quantum fields to create particles, and that the word `classical' simply refers to the lowest order approximation in quantum field theory. But, again, just like wave functions in non-relativistic quantum theory do not give rise to physical motion (of a particle or wave) in space and time -- without assumptions solving the measurement problem -- so quantum fields do not describe moving elementary particles in space with well-defined trajectories.

If we assume that this `quantum problem' could be properly dealt with, we would then have a sufficiently classical (or classicalized) inflaton field $\varphi$. The existence of such a field has been assumed  (as a mere postulate) in the investigation of various scalar field inflationary models since their inception  in the early 1980s. The background space for the inflationary multiverse is de Sitter space in which no matter is present (matter is only produced at the end of inflation inside the bubbles). Thus, the multiverse has -- as available  `material' to build up the reference frame from within -- only the
 inflaton field  $\varphi$. From this inflaton field one should construct some trajectories in order to satisfy the Weyl principle and thereby provide multiverse time (de Sitter $t$) a physical underpinning.\footnote{Greene (2012, p. 69) suggests that one may directly use the changing value of the $\varphi$ field as a clock (as measured by an ``inflaton-meter"). He apparently assumes that the $\varphi$ value is monotonically decreasing in de Sitter $t$. This idea is similar to the standard use, in FLRW cosmology, of matter density $\rho$, or the temperature $T$ of the background radiation, as a clock, see e.g. discussions in Rugh and Zinkernagel (2009). However, in our assessment, Greene's clock cannot in general trace the de Sitter $t$ ``time" parameter (and thus cannot provide a physical underpinning of it).  First because the (classical part of the) $\varphi$ field may not be homogeneous in $x$-space (as in Linde's chaotic model) -- and so the same $\varphi$ value (an `equal $\varphi$ hypersurface') becomes associated with different de Sitter $t$-values. And second because even an assumed homogeneously distributed $\varphi$ field will exhibit quantum fluctuations so that, again, the same $\varphi$ value gets mixed up with different $t$-values.}
One way of getting (a congruence of) non-crossing trajectories is to assume that the matter-energy content is in the form of a perfect fluid since this implies a well-defined four-velocity (and hence a direction for a trajectory) at each point of the spacetime manifold. As described e.g. by Krasinski (1997, p. 8) and Hobson et al. (2006, p. 432), a 4-velocity field of a perfect fluid can be constructed from (the gradient of) a scalar field.\footnote{The idea is to equate the energy momentum tensor of the perfect fluid form with the energy momentum tensor for the scalar field. This results in the 4-velocity 
$u_{\mu} = A \cdot \partial_{\mu} \varphi$ where $A = (\partial^{\nu} \varphi \; \partial_{\nu} \varphi)^{-1/2}$.} 
However, as we shall see in point 3 below, this may not be sufficient to satisfy the Weyl principle due to the fractal structure of the inflationary multiverse.

\subsubsection*{2. Could the trajectories be non-crossing? }

If the relevant substratum for the Weyl principle in the (inflationary) multiverse is the bubble- or pocket universes, there does indeed seem to be crossing of the trajectories.\footnote{If the substratum (the world-lines of which ``carry" the coordinates) is not formed by the bubble universes, but is rather to be found in the background de Sitter space with an inflaton field, then we are either back in the subsection above or proceed to the subsection below.} For instance, Garriga, Guth and Vilenkin (2006) note: 
\begin{quote}
A “bubble universe” nucleating in an eternally inflating false vacuum will experience, in the course of its expansion, collisions with an infinite number of other bubbles.
\end{quote}

Thus, bubble collisions do occur and so the Weyl principle is not satisfied at the level of bubbles. This problem appears to be aggravated by the observation that the inflationary multiverse seems to result in a fractal structure in which merging of different thermalized domains (bubble universes) occurs on all scales (see e.g. Guth 2007 and Vanchurin et al 2000).

\subsubsection*{3. Could crossing trajectories be ``averaged out"? }

Even if bubbles collide, and so trajectories cross, it may still be possible -- just as in the single universe case -- to devise an averaging procedure to ``smooth out" these crossings. However, this will be difficult in the realm of the inflationary multiverse since it appears to be fractal (Guth 2007, p. 6816). 
This means, as far as we can see, that there is no ``cut off" scale above which the implementation of averaging procedures will produce non-colliding
world-line trajectories out of bubbles (which collide below such
a scale).

If the Weyl substratum is to be constructed from the $\varphi$ field (outside the bubbles) the situation seems no better since these regions outside the bubbles likewise appear to form a fractal. This is suggested e.g. by the highly random and irregularly looking distribution of the scalar field(s) in fig. 20.2 in Linde (2004, p. 435) and explicitly stated in Vanchurin et al. (2000, p. 4): ``...these [inflating] regions [outside the bubbles] form a fractal of dimension $d<3$". Although this may not result in collisions
between trajectories constructed from the $\varphi$ field, it nevertheless seems to imply a problem concerning the averaging procedure. According to Guth (2007, p. 6816),
\begin{quotation}
\noindent           
One does have to think about the fractal structure if one wants
to understand the very large scale structure of the spacetime
produced by inflation.
\end{quotation}
We agree. But if one, indeed, thinks about exactly this, it appears that the fractal structure of the inflationary multiverse results in a far more complicated large scale spacetime structure than the highly symmetric Robertson-Walker spaces (which are isotropic and homogeneous) employed in simplified inflationary modeling. More fundamentally, in our assessment (and to be examined further), the Weyl principle appears not to be satisfied: According to this principle, the reference frame is built up from a space-time filling congruence of geodesics. This can at most be fulfilled in a coarse grained (averaging) sense. However, due to the self-similar fractal structure (of both the inflating and thermalized -- bubble -- regions) there is no possible coarse graining scale above which a spacetime filling congruence can be constructed (as there will be `holes' at all scales). If this is so, the physical foundation for a global de Sitter multiverse time appears insufficient.

\subsection{Extrapolating our cosmic time, proper time or time order back to an older bubble?}
\label{extrapolated}

If the Weyl principle does not hold in the multiverse, there will be no global time parameter which can be used to temporally order the bubble universes of {\em different} `branches' in figure 2. But it would seem that, even without a Weyl principle, it should still be possible to contemplate older structures than our universe by focusing on a {\em single} (our own) `branch' in the figure. Indeed, if we accept the idea that one bubble universe can somehow causally give rise to another, then it appears possible to consider other bubble universes (within our own `causal branch') which predate our universe. Nevertheless, as we shall indicate below, to contemplate this possibility may be far from straightforward. 

One way to address the causal past of our universe would be if we could extrapolate our `local' cosmic time concept further back than our (local) beginning. Now, if this beginning is taken to be (arbitrarily close to) an initial singularity or, alternatively, that it is located in some `epoch' described by quantum gravity such a proposal seems hopeless or, at best, highly speculative (see also Zinkernagel 2008). 
Indeed, most cosmologists would agree that there is no (known) sensible time concept ``before" the Planck time ($\sim 10^{-43} s$) and so no clear meaning can be ascribed to instants earlier than that.

However, if the beginning of our universe occurs -- as assumed in inflationary multiverse models -- at the beginning of the inflationary phase, then there may be no need to extrapolate time either through a singularity or through a quantum gravity epoch. Indeed, as long as some causal structure can be maintained (light cones should not tilt more than $45^{\circ}$), then it may be sensible to speak of the past of any event. Thus, one may perhaps speculate, for instance, that before the beginning of inflation at, say $10^{-35} s$, the universe no longer gets denser and hotter (as in standard cosmology) but rather expands into a previous bubble universe. In fact, such a suggestion may work even if the `local' (in our universe) Weyl principle is not satisfied in the inflationary epoch. For even if there is no cosmic time (no Weyl principle) it could still be possible to ask about the past of any event -- for instance, the past of the onset of inflation. Specifically, we can address the past of an event by extrapolating backwards proper time along a world-line which ends in the event. Such a possibility appears to be implied when Tegmark (2005, p. 49) remarks  (after stating, as we saw Garriga, Guth and Vilenkin do above, that geodesics cross after thermalization within a bubble):

\begin{quote}
When we discuss $t$ [time] for a particle in the present epoch, the rigorously inclined reader can simply take this to mean its proper time, since
this provides a well-defined ordering even after geodesic crossing. [our inserts]
\end{quote}

For this to be made into a workable suggestion for contemplating earlier bubbles than our own, it must be possible to identify (or, at least, to speculate) a particle world-line along which proper time can be extrapolated backwards.\footnote{From our relationist point of view -- in which time is necessarily related to physical processes (Rugh and Zinkernagel 2009) -- the time-like curves can only be identified (they only have a physical basis) if the motion of objects or test particles along these curves is at least in principle realizable from the available physics.}
 In particular, photons -- or other massless particles -- alone will not be sufficient as they have no past (i.e. their proper time is zero).\footnote{Within the framework of general relativity the notion of ``causal order" depends on the construction of ``backwards light cones" based on the existence of time-like or null-like curves (see e.g. Hawking and Ellis 1973, section 6 ``Causal structure") -- and therefore on the notion of (possible) classical particle or light-signal trajectories. The latter is insufficient to establish a chronological ordering of bubbles since -- if only light is present -- causal influences are instantaneous (again, photons have no past).}  Note that proper time along a specific world-line will give a quantitative measure of time differences between events. But since we are here only interested in the notion of {\em earlier} bubbles, a time (or chronological) order will be sufficient. Thus, the existence of {\em any} time-like curve (on which we can address proper times $\tau < \tau_0$, where $\tau_0$ is the beginning of our bubble) will suffice. 

In the inflationary scenario, the relevant candidate for a particle world-line (a time-like curve) will have to come from the $\varphi$ field. However, as discussed in section 2.1. and 3.1.1, there is a `quantum problem'  in constructing sensible notions of particle world-lines and classical trajectories from the inflaton field. In particular, at the supposed `birth' of a new bubble universe, the inflaton field is strongly quantum: Quantum fluctuations with amplitudes (within a factor of 10) of the order of the Planck scale are necessary to reset or lift the scalar field back to a value where a new bubble is born and becomes dominated by inflation (see e.g. Linde 2004, sect. 4).\footnote{Whereas Linde (2004) mostly discusses chaotic inflation, the quantum problem also shows up in the multiverse model based on the ``new inflation" scenario: It is hinted e.g. in Vilenkin (2004) that within new inflation, the scalar field is dominated by its quantum behavior when new bubble universes form (near the maximum of the inflaton potential).} Thus, at the `birth' of a new bubble universe, the $\varphi$ field is nowhere close to being a classical field on top of which we have small quantum fluctuations. Rather, it is entirely dominated by Planck scale quantum fluctuations.

It is therefore unclear to us how one would go about constructing any individual classical particle world-line from the inflationary scalar field $\varphi$ in a regime where its quantum behaviour is dominant.  But if such world-lines 
(classical trajectories) cannot be constructed from the underlying physics (the $\varphi$ field), it seems, in our assessment, that the very conditions for speaking about the past of an event in general relativity are not fulfilled. We therefore tentatively conclude that this proper time, or time order, route to contemplating earlier patches or bubbles (within a given branch of bubbles) in the multiverse seems problematic.\footnote{Guth (2007, p. 6822) reports a theorem according to which eternal inflation is not past-eternal (i.e. there must be a beginning of the inflating multiverse even though inflation always continues somewhere). This theorem focuses on the idea of a time-like (or null-like) geodesic which is, locally, extracted backwards to an ultimate (for the multiverse as a whole) big bang. The theorem seems to rest on the idea of a well-defined `local' congruence (of massive test particle world-lines) intersecting the geodesic. We would, again, object that the definitions of both the geodesic trajectory and the congruence are suspect if the underlying theory is of a quantum nature.}

\section{Outlook}

In this note we have argued that it is very difficult to construct a global multiverse
time parameter (as suggested e.g. by Linde and Guth) which would give a temporal ordering of different branches in the inflationary model of the multiverse (cf. figure 2). We have also indicated that it is not straightforward to maintain even a concept of time order within a given branch of bubbles since, at the birth of a bubble, the physics is entirely dominated by quantum fluctuations. This means that there is no possibility to construct classical trajectories (from the inflaton field) on which the causal and temporal structure in general relativity is based.
Thus, it is difficult to provide a physical underpinning of what one could mean by saying that some other bubble universe predates our own.

Our discussion above applies only to the restricted class of
(inflationary) multiverse models considered. As noted, these models appear to be the most elaborated versions of the multiverse -- in particular in terms of contemplated spatio-temporal structure (e.g. the notion of a background de Sitter space). In any case, it seems to us that it might even be more problematic to think of patches or bubbles `older' than ours if we consider more radical versions of the multiverse (for instance those contemplated in Tegmark's (2004) level III-IV). Such versions may include the notion of completely disconnected regions and/or fundamentally different physical laws in the different bubbles. This may well undermine (1) the causal structure needed to define the past light-cone of an event and, in particular, the idea of extrapolating proper time backwards to an earlier bubble; and (2) the possibility of comparing the time concepts of -- and thus temporally order -- different bubbles (e.g. since, as discussed in Rugh and Zinkernagel 2009, time is implicitly defined by laws). None of this means that there could not be ways to contemplate a multiverse older than our universe. But we would at least recommend that multiverse model builders ought to be clear about what time concept they use.

\section*{Acknowledgements}
We would like to thank Robert Brandenberger, George Ellis, Holger Bech Nielsen and an anonymous referee for helpful comments. We gratefully acknowledge the hospitality provided by the Niels Bohr Archive under the auspices of Finn Aaserud. HZ thanks the Spanish Ministry of Science and Innovation (Project FFI2011-29834-C03-02) and the Spanish Ministry of Education (Project PR2011-0150) for financial support.

\small

\section*{References}

\noindent
Carr B (ed) (2007) Universe or Multiverse? Cambridge: Cambridge University Press.

\noindent
Carr B, Ellis GFR (2008) Universe or multiverse? {\em Astronomy \& Geophysics} 49(2): 2.29--2.37.

\noindent
Ellis GFR (1999) 83 years of general relativity and cosmology: progress and problems. {\em Classical and Quantum Gravity} 16: A37--A75.

\noindent
Greene B. (2012) {\em The Hidden Reality. Parallel Universes and the Deep
Laws of the Cosmos}. London: Penguin Books.

\noindent
Guth AH (2007) Eternal inflation and its implications. {\em
Journal of Physics A: Mathematical and Theoretical} 40: 6811--6826.

\noindent
Hawking SW (1989) The edge of spacetime. In {\em The New Physics}, ed. Paul Davies, 61-69. Cambridge: Cambridge University Press..

\noindent
Hawking SW, Ellis GFR (1973) {\em The large scale structure of space-time}. 
Cambridge: Cambridge University Press.

\noindent
Hobson MP, Efstathiou GP, Lasenby AN (2006) {\em General Relativity}. Cambridge: Cambridge University Press.

\noindent
Krasinski A (1997) {\em Inhomogeneous Cosmological Models}. Cambridge: Cambridge University Press.

\noindent
Labini FS, Vasilyev NL, Pietronero L, Baryshev, YV (2009) Absence of self-averaging
and of homogeneity in the large-scale galaxy distribution. {\em Europhysics Letters} 86: 1--6.

\noindent
Linde A (1998) The Self-Reproducing Inflationary Universe. {\em Scientific American}  9(20): 98--104.

\noindent
Linde A (2004) Inflation, quantum cosmology, and the anthropic principle. In {\em Science and
Ultimate Reality}, ed. John D Barrow, Paul CW Davies and Charles L Harper, 426-458. Cambridge: Cambridge University Press. 

\noindent
Narlikar J (2002) {\em An introduction to cosmology} (3rd edition). Cambridge: Cambridge University Press.

\noindent
Rugh SE, Zinkernagel H (2005) Cosmology and the Meaning of Time, 76pp. Distributed manuscript.

\noindent
Rugh SE and Zinkernagel H (2009) On the physical basis of cosmic time.
{\em Studies in History and Philosophy of Modern Physics} 40: 1--19.

\noindent
Rugh SE and Zinkernagel H (2011) Weyl's principle, Cosmic Time and Quantum
Fundamentalism. In {\em Explanation, Prediction and Confirmation. The Philosophy of
Science in a European Perspective}, ed. Dennis Dieks et al., 411-424. Berlin: Springer Verlag.

\noindent
Rugh SE and Zinkernagel H (2013) A critical study of time in the universe and the multiverse. In preparation.

\noindent
Sudarsky D (2011) Shortcomings in the Understanding of Why Cosmological Perturbations Look Classical. {\em International Journal of Modern Physics D} 20: 509--552.

\noindent
Tegmark M (2004) Parallel universes. In {\em Science and
Ultimate Reality}, ed. John D Barrow, Paul CW Davies and Charles L Harper, 459-491. Cambridge: Cambridge University Press.  

\noindent
Tegmark M (2005) What does inflation really predict? {\em Journal of Cosmology and Astroparticle Physics} 4(1): 1--75. 

\noindent
Vanchurin V, Vilenkin A, Winitzki S (2000) Predictability crisis
in inflationary cosmology and its resolution. {\em Physical Review D} 61: 1--17.

\noindent
Vilenkin A (2004) Eternal inflation and chaotic terminology. Preprint. http://arxiv.org/abs/gr-qc/0409055v1.

\noindent
Zinkernagel H (2008) Did Time have a Beginning? {\em International
Studies in the Philosophy of Science} 22(3): 237--258.

\noindent
Zinkernagel H (2011). Some trends in the philosophy of physics. 
{\em Theoria} 26(2): 215--241.

\end{document}